\documentclass{article}
\usepackage{spconf,amsmath,amssymb,graphicx,xcolor,subfig,tabularx,hyperref,numprint}
\usepackage[OT1]{fontenc} 
\hypersetup{
    colorlinks=true,
    linkcolor=blue,
    filecolor=magenta,      
    urlcolor=blue,
    citecolor=violet,
    pdftitle={Adapting Frechet Audio Distance for Generative Music Evaluation},
    pdfpagemode=FullScreen,
}
\npdecimalsign{.}
\nprounddigits{2}

\usepackage{cite}
\usepackage{balance}
\usepackage{booktabs}
\usepackage{fancyhdr}

\title{Adapting Frechet Audio Distance for Generative Music Evaluation}

\twoauthors
  {Azalea Gui\sthanks{Work done while intern at Microsoft Research.}}
	{University of Toronto\\
	azalea@hydev.org}
  {Hannes Gamper, Sebastian Braun, Dimitra Emmanouilidou}
	{Microsoft Research Redmond\\
	\{first.last\}@microsoft.com}

\begin{document}
\ninept

\fancypagestyle{copyright}{\fancyhf{}\renewcommand{\headrulewidth}{0pt}\fancyfoot[L]{\small \copyright 20XX IEEE. Personal use of this material is permitted. Permission from IEEE must be obtained for all other uses, in any current or future media, including reprinting/republishing this material for advertising or promotional purposes, creating new collective works, for resale or redistribution to servers or lists, or reuse of any copyrighted component of this work in other works.”}}

\thispagestyle{copyright}

\maketitle 

\begin{abstract}
The growing popularity of generative music models 
underlines the need for perceptually relevant, objective music quality metrics. 
The Frechet Audio Distance (FAD) is commonly used for this purpose even though its correlation with perceptual quality is understudied. 
We show that FAD performance may be hampered by sample size bias, poor choice of audio embeddings, or the use of biased or low-quality reference sets.
We propose reducing sample size bias by extrapolating scores towards an infinite sample size. 
Through comparisons with MusicCaps labels and a listening test we identify audio embeddings and music reference sets that yield FAD scores well-correlated with acoustic and musical quality. 
Our results suggest that per-song FAD can be useful to identify outlier samples and predict perceptual quality for a range of music sets and generative models.
Finally, we release a toolkit that allows adapting FAD for generative music evaluation.

\end{abstract}

\begin{keywords}
  Generative model, Music/Audio generation, Frechet Audio Distance, Evaluation Metrics, Quality Assessment
\end{keywords}

\section{Introduction}
\label{sec:introduction}
Advancements in machine learning have accelerated progress in automated music creation. Recently, there has been growing interest in models that directly output music as opposed to a symbolic representation like MIDI~\cite{copet_simple_2023,agostinelli_musiclm_2023,yang_diffsound_2023,huang_noise2music_2023,liu_audioldm_2023,mariani_multi-source_2023,goel_its_2022}. However, evaluating the perceptual quality of music generated by these models remains a challenge.

Various metrics have been proposed for automatic music evaluation. Some metrics prioritize prompt adherence, including the CLAP score~\cite{elizalde_clap_2022}, MuLan Cycle Consistency~\cite{huang_mulan_2022}, and Kullback-Leibler (KL) divergence over audio classifiers. Other metrics emphasize latent proximity toward a ``studio-quality'' reference set, such as the Inception Score and the increasingly popular Frechet Audio Distance (FAD)~\cite{kilgour_frechet_2019}. FAD was adapted from the Frechet Inception Distance (FID) used in the image domain. A common approach to calculate FAD is to use the MusicCaps dataset~\cite{agostinelli_musiclm_2023} as a reference set and extract audio embeddings using the VGGish model~\cite{hershey_cnn_2017}, yielding a single FAD score for a set of generated music samples.

Despite being commonly used, 
prior work suggests that existing objective quality metrics including FAD fail to reliably predict the perceptual quality of generative music~\cite{vinay_evaluating_2022}.
Furthermore, FAD scores published in the literature are often obtained using a variety of sample sizes, reference music sets, and embeddings, all of which may affect FAD performance as a perceptual metric and make comparisons between reported results difficult or impossible.

Here, we examine limitations of FAD and propose improvements for evaluating generative music. 
Our main contributions are:
\begin{itemize}
    \item introducing FAD$_\infty$ (based on FID$_\infty$~\cite{chong_effectively_2020}) to reduce sample size bias;
    \item evaluating per-song FAD to identify outliers in a sample set; 
    \item evaluating two reference sets sourced from public datasets with permissive licensing and a variety of state-of-the-art audio embedding models objectively and subjectively to obtain FAD scores that correlate well with perceptual quality.
\end{itemize}

We release an FAD toolkit geared towards music evaluation.\footnote{Code available at \href{https://github.com/microsoft/fadtk}{github.com/microsoft/fadtk}}

\section{Background}
\label{sec:background}
Kilgour et al.\ proposed adapting the Frechet Inception distance (FID) for evaluating music enhancement methods~\cite{kilgour_frechet_2019}. Similarly to the original FID, the Frechet audio distance (FAD) has since become a commonly used objective metric for evaluating generative models. 
The FAD is computed by comparing a set of audio samples to a reference set in terms of their respective distributions in an embedding space. Given a multinormal fit to the distribution of audio embedding features with means $\mu_r$ and $\mu_t$ and covariance matrices $\Sigma_r$ and $\Sigma_t$  for the reference and test set, respectively, the FAD is~\cite{kilgour_frechet_2019}
\begin{equation}
    \mathrm{FAD} = \| \mu_r - \mu_t \|^2 + tr\left( \Sigma_r + \Sigma_t - 2\sqrt{\Sigma_r\Sigma_t}\right),
\label{eq:FAD}
\end{equation}
where $tr(\cdot)$ is the matrix trace. 
While FAD is often used as a proxy for perceptual quality~\cite{agostinelli_musiclm_2023,copet_simple_2023}, the underlying assumption given (\ref{eq:FAD}) is that the reference set is of high quality, that the audio embeddings capture features related to quality, that the embedding distribution can be approximated by a multinormal fit, and that the resulting single FAD score for the entire test set is a meaningful metric of model performance. In the following, we test some of these assumptions for several reference sets and audio embedding models and propose improvements to address some limitations of FAD. 

\section{FAD for music evaluation}
\label{sec:proposed-method}
\subsection{Reference sets for quality estimation}
\label{sec:refsets}

Since FAD is comparison-based, a lower distance only infers high quality if the reference set is of high quality. 
In the literature, a commonly used reference set for FAD calculation is MusicCaps which consists of 10~s music-related segments from Youtube videos~\cite{agostinelli_musiclm_2023}. However, MusicCaps captions indicate that a large portion of the samples are labeled ``low-quality'' by the expert raters (see Table~\ref{tab:aq-mq}).
Furthermore, upon manual inspection, some MusicCaps samples are not music but music-related tutorials, for example.

We propose two new reference sets based on publicly available music datasets. 
\emph{FMA-Pop} consists of 4230 songs from the Free Music Archive (FMA)~\cite{defferrard_fma_2017}. 
We selected the 30 most popular songs in 163 genres as ranked by FMA's ``pop'' label. We excluded speech-only genres like "spoken" and "poetry", and non-descriptive genres like "soundtrack" or "composed music". 
\emph{MusCC} is a combination of 49 songs from the musdb18 dataset released under a Creative Commons license~\cite{musdb18} and all 240 songs published in the MoisesDB dataset under a Non-Commercial Research Community license~\cite{pereira2023moisesdb}. 
MusCC consists of studio-quality, professionally mixed music while FMA-Pop consists of a range of recordings and large genre diversity. 

\subsection{Audio embedding models for quality estimation}
\label{sec:embeddings}

FAD is computed on audio embeddings. However, embedding models trained with different architectures, losses, and data may capture different aspects of audio. For the Frechet Inception Distance (FID), studies found that the top ImageNet class highly influences FID scores even though it is not quality-related~\cite{kynkaanniemi_role_2023}, and that low-probability outliers over the Inception embedding are not necessarily low quality~\cite{betzalel_study_2022}. In the music domain, there appears to be no consensus on which embedding might be preferred for capturing music quality. A common choice is VGGish but other embedding models such as OpenL3~\cite{cramer_openl3_2019}, TRILL~\cite{shor_trill_2020}, and MuLan~\cite{agostinelli_musiclm_2023} are also used to evaluate various generative models.

To assess the impact of the audio embedding on the perceptual aspects captured by the FAD we evaluate a set of embeddings covering a variety of training objectives: 
VGGish (audio classifier), CLAP~\cite{clap2} and LAION CLAP~\cite{laionclap2023} (joint audio--text embeddings), MERT~\cite{li_mert_2023} (self-supervised audio embedding), CDPAM~\cite{manocha_cdpam_2021} (audio similarity), and EnCodec~\cite{defossez_high_2022} and Descript Audio Codec (DAC)~\cite{kumar_high-fidelity_2023} (low-rate audio codecs). 
For CLAP, we use model ``sep23''. For LAION CLAP, we use ``630k-audioset-best'' (L-CLAP aud) and ``music\_audioset\_epoch\_15\_esc\_90.14'' (L-CLAP mus). For MERT, we use the last transformer layer (MERT) or the \emph{n}th layer (MERT L\emph{n}). For CDPAM we use the models ``acoustic'' (CDPAM ac) and ``content'' (CDPAM cont). 
For EnCodec, EnCodec 48k, and DAC, the embedding is extracted from the final layer of the encoder network \emph{prior} to quantization.
Table~\ref{tab:embedding_details} summarizes the embeddings. 

\begin{table}
    \setlength{\tabcolsep}{5pt}
    \centering
    \caption{Evaluated embedding models.}
    \begin{tabular}{l l l l l l}
    \toprule
        {Embedding model} & {Input channels} & {Size} & {Context} & {Hop} \\
    \cmidrule(lr){1-5}
    VGGish~\cite{hershey_cnn_2017} & 1 (16~kHz) & 128 & .96~s & .96~s \\
    CLAP~\cite{clap2} & 1 (44.1 kHz) & 1024 & 7~s & 1~s \\
    L-CLAP~\cite{laionclap2023} & 1 (48 kHz) & 512 & 10~s & 1~s \\
    MERT~\cite{li_mert_2023} & 1 (24 kHz) & 768 & 5s & .013~s \\
    CDPAM~\cite{manocha_cdpam_2021} & 1 (22.05 kHz) & 512 & 5s & 1~s \\
    EnCodec~\cite{defossez_high_2022} & 1 (24 kHz) & 128 & - & .013~s \\
    EnCodec 48k~\cite{defossez_high_2022} & 2 (48 kHz) & 128 & 1~s & .99~s \\
    DAC~\cite{kumar_high-fidelity_2023} & 2 (44.1 kHz) & 1024 & 5~s & .012~s \\
    \bottomrule
    \end{tabular}
    \label{tab:embedding_details}
\end{table}

\subsection{Addressing sample size bias}
\label{sec:sample_bias}

\begin{figure}
    \centering
    \includegraphics{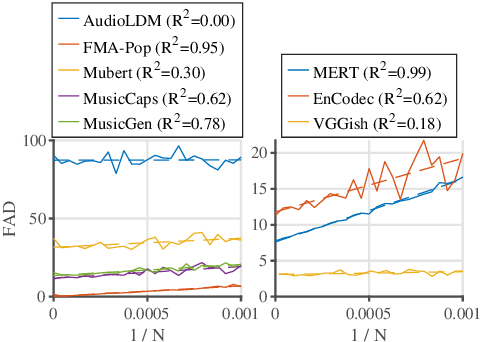}
    \caption{FAD score bias for EnCodec  and various testsets (left) and for various embeddings and MusicCaps (right), both with FMA-Pop as reference. For details on Mubert and MusicGen see section~\ref{sec:listening_test}.}
    \label{fig:inf-show-bias}
\end{figure}

Chong and Forsyth show that the Frechet Inception Distance (FID) decreases as the sample size $N$ increases; the bias coefficient (or rate of change) differs between evaluated models; and that FID converges as $N$ approaches infinity~\cite{chong_effectively_2020}. 
As seen in Fig.~\ref{fig:inf-show-bias}, sample size and model biases are also present in FAD, i.e., FAD generally decreases with increasing $N$ but at different rates for different models (e.g., AudioLDM~\cite{liu_audioldm_2023} vs.\ Mubert~\cite{Mubert}) and embeddings (e.g., MERT vs.\ EnCodec). These biases may impact the perceptual relevance of FAD scores. However, prior work introducing generative music models including MusicLM~\cite{agostinelli_musiclm_2023}, Noise2Music~\cite{huang_noise2music_2023}, and MusicGen~\cite{copet_simple_2023} typically does not disclose the sample size or evaluation data. 

To correct sample size bias for FAD, we propose a method similar to FID$_\infty$~\cite{chong_effectively_2020}. 
Given a test set with $N$ audio embedding frames, 
we compute FAD scores 
for audio frames sampled randomly with replacement from the test set for different sample sizes 
and fit a linear regression model to the scores to extrapolate the unbiased FAD$_\infty$ estimate at N = $\infty$. As shown in Fig.~\ref{fig:inf-show-bias}, the linear regression fits the data reasonably well for various testsets and audio embeddings.

\section{Experimental evaluation}
\label{sec:experiment}
\subsection{Acoustic Sensitivity Test}
\label{sec:distortions}

\begin{figure}
    \centering
    \includegraphics{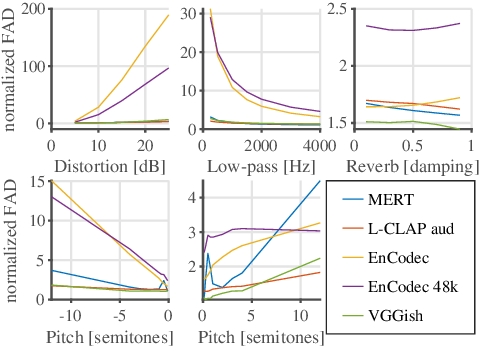}
    \caption{Impact of audio effects on FAD scores (normalized by FAD for unprocessed samples to show relative change).}
    \label{fig:distortion_test_results}
\end{figure}

Similarly to Kilgour et al.~\cite{kilgour_frechet_2019}, we 
evaluate the sensitivity of FAD, calculated using different embeddings, to acoustic effects and distortions artificially added to a random sample of 100 songs from FMA-Pop. 
Five effects are implemented using pedalboard~\cite{pedalboard}: distortion, low-pass filtering, reverberation, pitch down, and pitch up.
Figure~\ref{fig:distortion_test_results} illustrates the sensitivity of each embedding to the tested distortions. EnCodec and EnCodec-48k are particularly sensitive to distortion and low-pass filtering. All models seem somewhat sensitive to pitch shifting, while none seem overly sensitive to reverberation. This may indicate that the embeddings are sensitive to potential artifacts introduced by audio effects like pitch shifting, and that EnCodec penalizes acoustic degradation similar to distortion. 

\subsection{Predicting MusicCaps quality with per-song FAD}

We evaluate the FAD scores' effectiveness for capturing perceptual aspects related to  acoustic quality (AQ), i.e., audio recording quality in terms of noise and artifacts, and musical quality (MQ), i.e., quality of the musical composition and performance. 

We analyse MusicCaps captions using GPT-4 via the Azure OpenAI service to obtain subjective AQ and MQ labels ``high'', ``medium'', ``low'', or ``not mentioned'' for each song. As an example, the caption \emph{``The low quality recording features a ballad song that contains sustained strings, mellow piano melody and soft female vocal singing over it. It sounds sad and soulful, like something you would hear at Sunday services.''} is assigned the labels ``low'' for AQ and ``high'' for MQ. The AQ and MQ labels extracted from MusicCaps captions are summarised in Table~\ref{tab:aq-mq}. To balance the labels we aggregate AQ labels into the classes ``low'' and ``not low'', and MQ labels into the classes ``high'' and ``not high''.\footnote{GPT-4 MusicCaps labels available at \href{https://github.com/microsoft/fadtk}{github.com/microsoft/fadtk}}

\begin{table}
    \centering
    \caption{Acoustic (AQ) and musical quality (MQ) labels extracted from MusicCaps captions using GPT4.}
    \begin{tabular}{lcccc}
    \toprule
        & High & Medium & Low & N/A \\
        \cmidrule(lr){1-5}
        {Acoustic quality} & 1\% & 2\% & 39\% & 58\% \\
        {Musical quality} & 42\% & 9\% & 9\% & 41\% \\
        \bottomrule
    \end{tabular}
    \label{tab:aq-mq}
\end{table}

\begin{figure}
    \centering
    \includegraphics{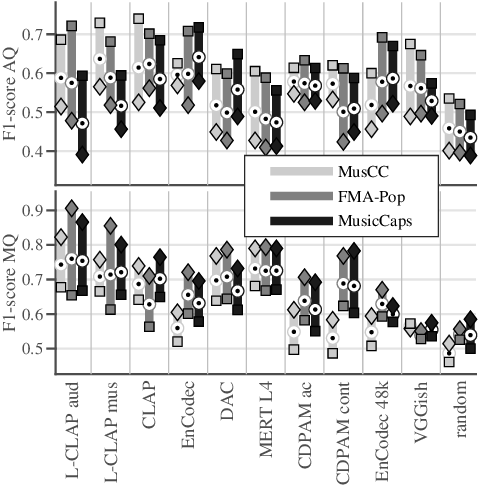}
    \caption{Precision (diamond), recall (square), and F1 score (circle) of FAD predicting MusicCaps acoustic (AQ, top) and musical quality (MQ, bottom) with FMA-Pop, MusCC, or MusicCaps as reference. The F1 score is the harmonic mean of precision and recall.}
    \label{fig:MusicCaps_FAD}
\end{figure}

For each audio embedding, we calculate an individual FAD score for every song in the MusicCaps dataset using FMA-Pop, MusCC, and MusicCaps itself as reference and  select the top and bottom 5\% (i.e., 276 songs) according to their FAD score. To test whether these FAD scores are perceptually meaningful, we assign the AQ label ``low'' to the songs with the highest FAD scores and the MQ label ``high'' to the songs with the lowest FAD scores and compare these FAD-predicted labels with the AQ and MQ labels extracted from the corresponding MusicCaps captions using GPT4. 

Figure~\ref{fig:MusicCaps_FAD} illustrates the performance of per-song FAD scores for predicting MusicCaps AQ and MQ labels in terms of precision, recall, and F1 score. CLAP and L-CLAP outperform other embeddings overall, perhaps indicating that the joint audio--text embeddings capture both acoustic and musical characteristics. However, they see a drop in performance for AQ prediction when using MusicCaps as a reference set, possibly due to the presence of low-quality samples in MusicCaps (cf.\ Table~\ref{tab:aq-mq}). 
EnCodec performs well for AQ prediction, in line with its high sensitivity to acoustic distortions (cf.\ Section~\ref{sec:distortions}).  
Among all MERT transformer layers, we found layer 4 to produce the best results, performing similarly to CLAP and L-CLAP for MQ prediction. This is in line with results reported by MERT authors indicating that earlier transformer layers may show perform better for certain music information retrieval tasks~\cite{li_mert_2023}.
VGGish predicts MQ labels only slightly better than chance, perhaps due to a lack of music-specific labels or data in training. 

\subsection{Identifying outliers with per-song FAD}
To further gauge how the choice of audio embedding and reference set might impact which acoustic and musical characteristics are captured by FAD we check the MusicCaps musical excerpts that receive the five highest and five lowest per-song FAD for various combinations of audio embedding and reference set.\footnote{Examples available at \href{https://fadtk.hydev.org/}{fadtk.hydev.org}} 
For MERT L4, MusicCaps samples that consist of synthetic tones with little or no temporal or spectral variation consistently receive the highest FAD scores, irrespective of the reference set used, indicating per-song FAD with MERT may be useful to identify samples of low musical information. Among the excerpts with the highest FAD scores using L-CLAP aud are non-musical field recordings, a speech-only sample, and monophonic synthesizer sounds. The excerpts with the lowest FAD scores for L-CLAP aud consist of studio-quality music, but the genre diversity seems to depend on the reference set. When using MusCC, all five excerpts with the lowest FAD score fall within the Western music genres Pop and Rock (consistent with the dominant genres in MusCC), whereas using FMA-Pop the five excerpts span a variety of genres, including Jazz and Latin music. This suggests that genre-imbalance in the reference set may bias FAD scores.
For EnCodec, the highest FAD scores are assigned to samples with heavy distortion or clipping. This is again in line with results from the acoustic sensitivity test (see Section~\ref{sec:distortions}). 
In the case of VGGish embeddings with MusicCaps as a reference, i.e., a common choice for FAD calculation in the generative music literature, four out of five samples with the highest FAD scores consist of bagpipe music. Interestingly, Kong et al.\ report that for AudioSet classification (i.e., the same training objective as VGGish) the class ``bagpipes'' achieves the highest average precision out of all 527 classes~\cite{kong2020panns}. It is possible that particularly recognizable classes may produce unique embedding activations and thus result in high FAD scores. 
These anecdotal findings suggest that per-song FAD scores may be a useful tool to find outliers (e.g., heavily distorted, noisy, or non-musical samples) or to check whether the chosen audio embedding and reference set may introduce a bias with respect to musical genre or instruments. 
\subsection{Subjective evaluation}
\label{sec:listening_test}

\begin{figure}
    \centering
    \includegraphics{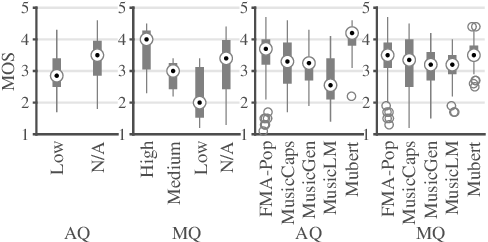}
    \caption{Mean opinion score (MOS) of acoustic quality (AQ) and musical quality (MQ) for MusicCaps labels extracted by GPT4 (left two panels) and for all datasets in the listening test (right two panels). Note: The left two panels consist of only the MusicCaps samples within the listening test subset and not the entire MusicCaps set.}
    \label{fig:mcaps_MOS}
\end{figure}

To evaluate the performance of per-song FAD scores calculated using various audio embeddings and reference sets for predicting subjective quality of generative music we conducted a listening test. The test dataset consisted of 300 10~s excerpts from five sources: 100 random songs from FMA-Pop, 50 random songs from MusicCaps~\cite{agostinelli_musiclm_2023}, and 50 songs each generated by Mubert~\cite{Mubert}, 
MusicLM~\cite{agostinelli_musiclm_2023}, and MusicGen (Large)~\cite{copet_simple_2023}. 
The prompts for the musical excerpts generated using Mubert, MusicLM, and MusicGen were derived from the 50 MusicCaps captions via GPT-4 with a one-shot example.\footnote{GPT-4 prompt and results available at \href{https://github.com/microsoft/fadtk}{github.com/microsoft/fadtk}}
During the test, participants were asked to rate each excerpt in terms of acoustic quality (AQ) and musical quality (MQ) on a five-point Likert scale. The order of excerpts was fully randomized. We recruited 10 participants within Microsoft. Each participant completed the test in approximately one hour using a pair of AKG K271 MkII headphones. All participants received a \$25 gift card. The study design was approved by Microsoft's Ethics Review Board. 

Figure~\ref{fig:mcaps_MOS} shows the Mean Opinion Score (MOS) distribution for all testsets in the listening test. As can be seen , a ``low'' AQ rating assigned by GPT-4 indeed corresponds to a lower MOS compared to ``N/A' (i.e., no label assigned). Similarly, ``high'', ``medium'', and ``low'' MQ labels assigned by GPT-4 correlate well with MOS results. Among the five testsets, Mubert achieves the highest median MOS for AQ, followed by FMA-Pop. 
For MQ, the differences between the testsets are less pronounced. Note that listeners were not asked to rate prompt adherence. 

\begin{figure}
    \centering
    \includegraphics{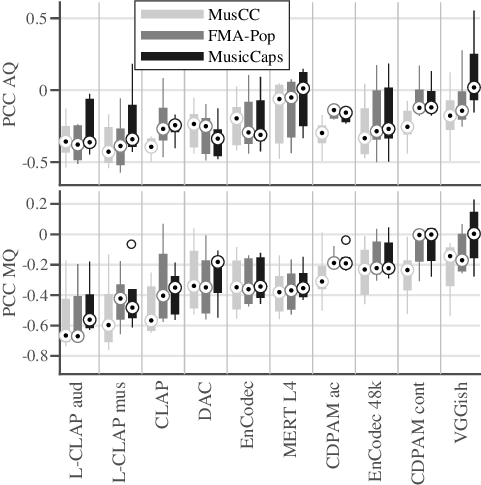}
    \caption{Pearson correlation coefficient (PCC) between FAD and listening test MOS for all tested embeddings and reference datasets, for acoustic (AQ, top) and musical quality (MQ, bottom). Each box summarizes the five PCCs of the five testsets; lower PCC is better. Each box on the boxplot shows five PCCs associated with the five testsets.}
    \label{fig:FAD_PCC}
\end{figure}

Figure~\ref{fig:FAD_PCC} summarizes the Pearson correlation coefficients (PCCs) between per-song FAD and MOS. A PCC is calculated separately for each of the five testsets to measure per-song FAD performance for predicting MOS within each testset, and to account for potential model or dataset bias (cf.\ Section~\ref{sec:sample_bias}). Similarly to the results for MusicCaps label prediction, CLAP and L-CLAP outperform other embeddings for AQ and MQ MOS prediction. DAC and EnCodec perform similarly. In line with MusicCaps results, MERT L4 performs poorly for AQ MOS prediction and fares better for MQ. Finally, VGGish does not correlate well with either AQ or MQ MOS. When comparing the three reference sets used for FAD calculation, MusCC seems to boost MQ MOS prediction compared to MusicCaps especially for CLAP and L-CLAP models, whereas for AQ prediction the differences are less clear.  

\section{Conclusion}
\label{sec:conclusion}
Automated evaluation of generative music models remains a challenging problem. Despite its limitations, the Frechet Audio Distance (FAD) is a promising metric for rating the quality of generative music. 
We evaluated several state-of-the-art audio embeddings for FAD calculation as well as two new reference sets derived from public datasets.
Our experimental results are in line with prior work suggesting that sample size bias in FAD calculation exists but can be estimated and mitigated~\cite{chong_effectively_2020}, and that embeddings trained on a classification objective such as VGGish may not be optimal for FAD calculation~\cite{betzalel_study_2022,kynkaanniemi_role_2023}. Evaluations on subjective labels extracted from MusicCaps captions as well as a listening test indicate that a proper choice of reference dataset and audio embedding results in FAD scores that meaningfully correlate with subjective ratings of generative music in terms of acoustic and musical quality. Furthermore, the results suggest that FAD scores calculated for individual samples rather than an entire testset at once may prove useful for identifying outliers, including samples with severe artifacts, atypical musical characteristics, or non-music signals. We hope these results along with the resources published with our FAD toolkit support future work on generative music.

\balance
\bibliographystyle{IEEEbib_abbr}
\bibliography{refs,refs-gen,refs-evaluation,refs-features,refs-datasets} 

\end{document}